\documentclass[iop]{emulateapj}
\usepackage{subfigure}

\newcommand{\gcmc}{\ensuremath{\rm g\,cm^{-3}}}
\newcommand{\teff}{\ensuremath{T_{\rm eff}}}

\newcommand{\logg}{\ensuremath{\log{g}}}

\newcommand{\feh}{[Fe/H]}

\newcommand{\rsun}{\ensuremath{R_\sun} }
\newcommand{\msun}{\ensuremath{M_\sun} }

\newcommand{\rstar}{\ensuremath{R_\star}}
\newcommand{\mstar}{\ensuremath{M_\star}}

\newcommand{\rhostar}{\ensuremath{\rho_\star}}

\newcommand{\rpl}{\ensuremath{R_{\rm p}}}

\newcommand{\rprs}{\ensuremath{R_{\rm p}/R_\star}}

\newcommand{\adrs}{\ensuremath{a/R_\star}}

\newcommand{\rearth}{\ensuremath{R_{\earth}}}

\newcommand{\ikt}{{\it Kepler}}
\newcommand{\ik}{{\it Kepler~}}

\newcommand{\keppl}{Kepler-1649b}
\newcommand{\kepst}{Kepler-1649}
\newcommand{\koipl}{KOI-3138}

\newcommand{\steff}{\ensuremath{3240\pm61}}
\newcommand{\slogg}{\ensuremath{4.98\pm0.22}}
\newcommand{\sfeh}{\ensuremath{-0.15\pm0.11}}
\newcommand{\srad}{\ensuremath{0.252\pm0.039}} %from spectrum
\newcommand{\smass}{\ensuremath{0.219\pm0.022}} %from spectrum

\newcommand{\tdur}{\ensuremath{1.0357\pm0.0966 {\rm\ hours}}}
\newcommand{\tdep}{\ensuremath{1783\pm101\ {\rm ppm}}}
\newcommand{\per}{\ensuremath{8.689090\pm0.000024}}
\newcommand{\epoch}{\ensuremath{2454966.2348\pm0.0026}}
\newcommand{\impact}{\ensuremath{0.34^{+0.15}_{-0.34}}}
\newcommand{\rds}{\ensuremath{0.0391^{+0.0014}_{-0.0022}}}

\newcommand{\prad}{\ensuremath{1.08\pm0.15}}
\newcommand{\sinc}{\ensuremath{2.30\pm0.65}} %incident flux
\newcommand{\ads}{\ensuremath{60.6\pm8.1}}
\newcommand{\incl}{\ensuremath{89.57\pm0.32}}
\newcommand{\asemi}{\ensuremath{0.0514\pm0.0028}}

\shorttitle{An Exo-Venus In the Solar Neighborhood}
\shortauthors{Angelo et al.}
\begin{document}
%Hey look, we found lots of exoplanets.
%\title{Kepler-XX is not a Mars sized Planet in the Habitable Zone}
\title{Kepler-1649b: An Exo-Venus in the Solar Neighborhood}

\author{Isabel Angelo,\altaffilmark{1,2,3},Jason~F. Rowe,\altaffilmark{1,2,4,12}, Steve B. Howell\altaffilmark{2}, Elisa V. Quintana\altaffilmark{1,2}, Martin Still\altaffilmark{5}, Andrew W. Mann\altaffilmark{6}, Ben Burningham\altaffilmark{2,7}, Thomas Barclay\altaffilmark{2,5}, David R. Ciardi\altaffilmark{8},  Daniel Huber\altaffilmark{1,9,10}, Stephen R. Kane\altaffilmark{11}  \\
\email{isabelangelo@berkeley.edu}
} 

\altaffiltext{1}{SETI Institute, Mountain View, CA 94043}
\altaffiltext{2}{NASA Ames Research Center, Moffett Field, CA 94035, USA}
\altaffiltext{3}{Department of Astronomy, University of California, Berkeley, CA, 94720, USA}
\altaffiltext{4}{Institut de recherche sur les exoplan\`etes, iREx, D\'epartement de physique, Universit\'e de Montr\'eal, Montr\'eal, QC, H3C 3J7, Canada}
\altaffiltext{5}{Bay Area Environmental Research Institute, 625 2nd St. Ste 209 Petaluma, CA 94952, USA}
\altaffiltext{6}{Hubble Fellow, Department of Astronomy, The University of Texas at Austin, Austin, TX 78712, USA}
\altaffiltext{7}{Centre for Astrophysics Research, School of Physics, Astronomy and Mathematics, University of Hertfordshire, Hatfield AL10 9AB, UK}
\altaffiltext{8}{NASA Exoplanet Science Institute/Caltech, Pasadena, CA, USA}
\altaffiltext{9}{Sydney Institute for Astronomy (SIfA), School of Physics, University of Sydney, NSW 2006, Australia}
\altaffiltext{10}{Stellar Astrophysics Centre, Department of Physics and Astronomy, Aarhus University, Ny Munkegade 120, DK-8000 Aarhus C, Denmark}
\altaffiltext{11}{Department of Physics \& Astronomy, San Francisco State University, 1600 Holloway Avenue, San Francisco, CA 94132, USA}
\altaffiltext{12}{Department of Physics, Bishops University, 2600 College Street, Sherbrooke, QC, J1M 1Z7, Canada}

\begin{abstract}

The \ik mission has revealed that Earth-sized planets are common, and dozens have been discovered to orbit in or near their host star's habitable zone. A major focus in astronomy is to determine which of these exoplanets are likely to have Earth-like properties that are amenable to follow-up with both ground- and future space-based surveys, with an ultimate goal of probing their atmospheres to look for signs of life. Venus-like atmospheres will be of particular interest in these surveys. While Earth and Venus evolved to have similar sizes and densities, it remains unclear what factors led to the dramatic divergence of their atmospheres. Studying analogs to both Earth and Venus can thus shed light on the limits of habitability and the potential for life on known exoplanets. Here we present the discovery and confirmation of \keppl, an Earth-sized planet orbiting a nearby M5V star that receives incident flux at a level similar to that of Venus. We present our methods for characterizing the star, using a combination of PSF photometry, ground-based spectroscopy and imaging, to confirm the planetary nature of \keppl. Planets like \keppl\ will be prime candidates for atmospheric and habitability studies in the next generation of space missions.

\end{abstract}

% something about how Kepler can precisely measure the period etc., but it only measures the ratio of the size of the planet to its host star. Therefore it is crucial that the star be well characterized 

%\keywords{planets: KOI-3138b --- Facilities: \facility{The \ik Mission}.}

\section{Introduction}\label{sec:intro}

The \ik mission was designed to measure the frequency and sizes of extrasolar planets~\citep{Borucki2010a}, with a primary goal of detecting other Earth-sized planets that could potentially be habitable. In our Solar System, both Earth and Venus evolved to have comparable sizes and bulk densities, yet the evolution of their atmospheres diverged dramatically such that only Earth developed conditions conducive to the emergence of life. It remains unclear which aspects of the Earth's development were key in acquiring and maintaining a hospitable atmosphere. Finding and characterizing both Earth and Venus analogs around other stars could shed light on these differences. 

\ik has been successful in finding Earth-size planets in the habitable zones of their host stars \citep{Quintana2014,Torres15}. A super-Earth in a Venus-like orbit and dozens of small planet candidates that could potentially have Venus-like atmospheres have also been discovered \citep{Barclay2013b,Kane2013}. In this paper we confirm the planetary nature of \keppl\ (KOI-3138.01), an Earth-sized planet that receives flux from its host star that is comparable to that received by Venus. 

\kepst\ appears in the Kepler Input Catalog \citep[KIC,][]{Brown2011} as KIC 6444896 with a brightness of 17.131 magnitudes in the Kepler bandpass ($Kp$) and has a relatively high proper motion of 0.157$\arcsec$ yr$^{-1}$ \citep{Lepine2005}. It was not selected as a prime mission target \citep{Batalha2010} but was proposed as part of Cycle 2 of the NASA Guest Observer (GO) Program (GO20031) to search for gravitational lensing in the \ik field of view \citep{DiStefano2012}.  Through the GO program 1 year of long-cadence (30 minute) observations covering quarters Q6—-Q9 were collected. A transit with a period of 8.7 days was detected and the target was given the designation Kepler Object of Interest (KOI) 3138 \citep{Burke2014}. After the discovery and dispositioning of the transit event, the target was added to the prime exoplanet target list for Q12--Q17.

In the Q1-Q12 catalogue \citep{Rowe2015}, \keppl\ was noted as an interesting cool sub-Earth radius planet candidate in a 8.7-day period around a cool M-dwarf (\teff = 2703) based on broadband photometric colours \citep{Huber2014}. The fitted value of the mean-stellar density (\rhostar) of $70\pm^{25}_{42}$ \gcmc and short transit-duration, $1.04\pm0.10$ hours, as reported in the Q1-Q12 catalogue, were consistent with the cool dwarf characterization of the host star.

While \keppl\ shares a similar size and incident flux as Venus, it orbits a nearby (219 ly) M5V star that is about a quarter of the size and mass of our Sun. Estimates on the size of this planet evolved as better constraints on the star's properties were attained over a period of several years from its initial detection. Although \ik photometry provides the orbital period and the planet's size relative to its host star to high precision, characterizing the transiting exoplanet is typically more limited since its fundamental properties depend critically on properties of the host star. For the bulk of the \ik planet sample, these stellar properties are based on matching broadband photometric measurements to stellar evolution models with various choices for priors that may or may not account for observational biases \citep{Brown2011,Batalha2013,Huber2014}. Improving these stellar parameters with better diagnostics such as spectroscopy can help us learn about systematics that may skew our interpretation of the \ik sample. Additionally, follow-up high-resolution imaging of \ik planet candidates is crucial for constraining properties of the planet system and its stellar host(s), particularly because ~50\% of planet hosts are likely to be binary \citep{Horch2014}.

%A super-Earth in a Venus-like orbit has been confirmed (Barclay et al. 2014, Kane et al. 2013), as well as 

%\citep{Kane2014} than Earth or Mars. Constraining properties of Venus-Analogs like Kepler-XX will be important for upcoming missions like TESS and JWST, which are dedicated to the discovery and characterization of Earth-sized habitable zone planets.

%We then derive a set of properties of \kepst\ in \S\ref{sec:planetprop}. 

%%%\section{Initial estimates of Kepler-XX properties}\label{sec:initial}

%It can also improve the accuracy of efforts to study individual planet sizes and atmospheric properties in the interest of potential habitability \citep{Tarter2007}.

%There has also been much effort in studying individual planets to measure both masses and basic atmospheric properties in the interest of potential habitability \citep{Tarter2007}.  

The planet candidate in the \kepst\ (\koipl) system, was first listed in the \ik Q1--Q8 catalogue \citep{Burke2014} with a stellar radius of 0.6\rsun and a planet size of 4.6\rearth. Soon after the Q1-Q8 KOI release there was a substantial effort from the Kepler Stars Working Group to improve global estimates of the stellar parameters. Most stars in the Kepler sample, including \kepst, were re-fit using Dartmouth models \citep{Dotter2008}. This gave a smaller estimated stellar radius of 0.12 \rsun (Huber et al. 2014). Matching the new stellar parameters to improved transit models gave a planet with a radius of \rpl=0.57\rearth\ receiving an incident flux ($S$) of 0.47 relative to Earth around a host star with \teff\ =2703 K, as noted in the Q1-Q12 catalogue \citep{Rowe2015}. 

% add this to the burke sentence? check
%The temperature estimate in the Q1-Q8 catalogue was based on matching 2MASS J-K colour photometry to fixed metallicity stellar properties with the assumption that the star was on or near the Zero-Age-Main-Seqeunce. 

With the revised stellar parameters, the \kepst\ system appeared to host a Mars-sized planet around a cool, nearby M-dwarf with incident bolometric flux levels similar to Mars. These properties would make this planet the first `exoMars', and would add to the small sample of potentially rocky planets transiting in or near the habitable zone of nearby M-dwarfs.  We obtained followup spectroscopy and imaging to verify the planetary nature of the transiting planet and better constrain stellar properties. Such imaging would could also potentially detect companions or background stars that indicate a planet radius larger than that determined by the transit depth. We found that the host star is significantly larger and hotter than previously estimated. The radius and incident flux levels of the planet increased to 1.08 $\rearth$ and 2.30 {$S_\earth$}, respectively. Revised stellar properties that necessitate recharacterization of planet properties are not unique to the \kepst\ system, but rather a common occurence for \ik systems \citep{Everett2013,Huber2013,Gaidos2016}. This case provides an example of the caution needed when constraining a planet's size based on various star catalogs, and the value of followup observations to improve estimates of the host star properties. 
%such as those outlines in this section

%Such planets have shorter orbital periods and large planet-star ratios relative to Earth-Sun analogue which makes followup observations with additional facilities (e.g. JWST) to characterize a potential atmosphere amenable.
 %These properties made the planet
%interesting due to current interest in finding Earth-sized HZ planets transiting  
% From the thousands of planetary-candidates in the Kepler sample, \kepst\ did not strike the authors as particularly interesting. 

%In \S\ref{sec:initial} we discuss our early estimates of the planet properties and how they evolved with improved stellar constraints. 

%%Planets like \keppl\ will be valuable targets for future space missions that will probe atmospheric properties that can help us understand whether the atmospheres of Earth and Venus are typical, and also how Earth and Venus analogs differ when orbiting cool stars.

Herein we present our confirmation of \keppl\ using transit and stellar models combined with ground-based observations. In \S\ref{sec:spec} we present our ground-based spectroscopic followup observations and classification of \keppl. In \S\ref{sec:phot} we present our technique of point-spread-function (PSF) extracted photometry and light curve modeling to constrain the planet properties of \keppl.  Our validation of \keppl\ as a planet using ground-based followup imaging is presented in \S\ref{sec:validation}. Finally, we summarize our results and comment on relative comparisons to Venus in \S\ref{sec:discussion}.

\section{Spectroscopic Observations and Stellar Classification}\label{sec:spec}

%Our spectral observations of \kepst\ were made with the double beam spectrograph (DBSP) attached to the 200" Hale telescope. The dichroic filter D-55 was used to split light between the blue and red arms. The blue arm used a 1200 l mm$^{-1}$ grating providing R $\sim$ 7700 and covered 1500 \AA\ of spectrum. The red arm used a 1200 l mm$^{-1}$ grating providing R $\sim$ 10, 000 and covered 670 \AA. The slit width was set to 1$\arcsec$ and the usual procedures of observing spectrophotometric stars and arc lamps were adhered to. Red spectra were wavelength calibrated with a HeNeAr lamp while the blue arm used a FeAr lamp. The night was clear and provided stable seeing near 1$\arcsec$. 

%The spectra have a resolution of 0.82 \AA/pixel. 

%reduced with standard IRAF {\it twodspec} IRAF packages.

\begin{figure}
\begin{center}
\includegraphics[width=0.45\textwidth]{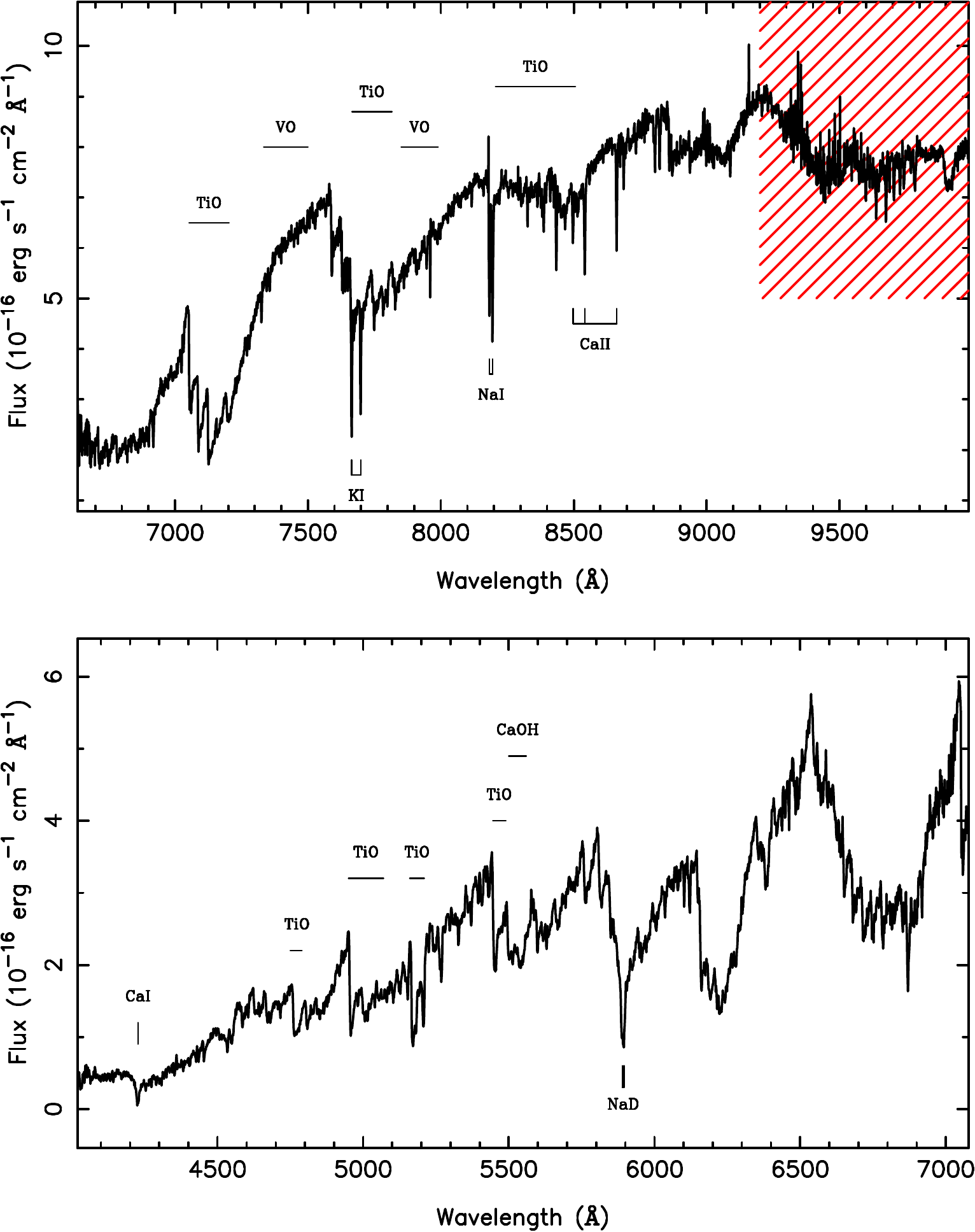}
\end{center}
\caption[Spectrum]{Spectrum of \kepst\ collected using the Double-Beam spectrograph on the 200$''$ Hale telescope at Palomar Observatory. The lower panel shows the blue beam and the upper panel the red beam. The spectrum is consistent with a classification of a mid-M dwarf star. The red hatched region is compromised because it is dominated by telluric lines and was not used in our analysis.}
\label{fig:spectrum}
\end{figure}

Spectroscopic observations of \kepst\ were made on 11 Feb 2015 with the Double-Beam spectrograph attached to the 200$''$ Hale reflector at the Mount Palomar Observatory. The dichroic filter D-68 was used to split light between the blue and red arms near 7000 \AA. The blue arm used a 1200 lines mm$^{-1}$ grating providing R $\sim$ 7700 and covered $\sim$2500  \AA\ of spectrum, 4200--7000 \AA. The red arm also used a 1200 lines mm$^{-1}$ grating providing R $\sim$ 9,000 and covered $\sim$2500 \AA\ of spectrum, 7000--9500 \AA. The spectra have a dispersion of approximately 0.82 \AA/pixel across the bandpass. The slit width was set to 1$''$, the integration time was 300 sec and the usual procedures of observing spectrophotometric stars and arc lamps were adhered to. Red spectra were wavelength calibrated with a HeNeAr lamp while the blue arm used a FeAr lamp. The nights were clear and provided stable seeing near 1$''$. Data reduction was done using IRAF two- and one-dimensional routines for spectroscopic data and produced a final one-dimensional spectrum for each observation, as shown in Figure \ref{fig:spectrum}. From analysis of these spectra, we were able to derive the effective temperature (\teff\ ), stellar radius (\rstar\ ), and stellar metallicity([Fe/H]) and stellar gravity (log$g$), as described in the following paragraphs. Our results are shown in Table 1.

\begin{deluxetable}{lcc}
\tabletypesize{\scriptsize}
\tablewidth{0pc}
\tablenum{1}
\tablecaption{System Parameters for \kepst \label{tab:parameters}}
\tablehead{\colhead{Parameter}	& 
\colhead{Value} 		& 
\colhead{Notes}}
\startdata
\sidehead{\em Transit and Orbital Parameters}
Orbital period $P$ (d) & \per & A \\ Midtransit time $E$ (HJD) &
\epoch & A \\ Scaled semimajor axis $a/\rstar$ & \ads & A
\\ Scaled planet radius \rpl/\rstar & \rds & A \\ Impact
parameter $b \equiv a \cos{i}/\rstar$ & \impact & A \\ Orbital
inclination $i$ (deg) & \incl & A \\
%Orbital eccentricity $e$ & $< 0.024$ & A,B,G \\/Users/rowe/Dropbox/Apps/TeX Writer/koi3138/koi3138.tex
\sidehead{\em Derived stellar parameters}
Effective temperature \teff\ (K)                & \steff        & B     \\
Spectroscopic gravity \logg\ (cgs)              & \slogg        & B     \\
Metallicity \feh                                & \sfeh         & B     \\
%\sidehead{\em Derived stellar parameters}
Mass \mstar (\msun)                             & \smass        & C   \\
Radius \rstar (\rsun)                           & \srad        & C   \\
%Surface gravity \loggstar\ (cgs)                & \koicurYYlogg         & C,D   \\
%Luminosity \lstar\ (\lsun)                      & \koicurYYlum          & C,D   \\
%Absolute V magnitude $M_V$ (mag)               & \koicurYYmv           & D     \\
%Age (Gyr)                                       & \koicurYYage          & C,D   \\
%Distance (pc)                                  & \koicurXdist          & D     \\
\sidehead{\em Planetary parameters}
%Mass \mpl\ (\mjup)                              & \koicurPPm            & A,B,C,D       \\
Radius \rpl\ (\rearth, equatorial)                & \prad            & A,B,C       \\
%Density \rhopl\ (\gcmc)                         & \koicurPPrho          & A,B,C,D       \\
%Surface gravity \loggpl\ (cgs)                  & \koicurPPlogg         & A,B,C,D       \\
Orbital semimajor axis $a$ (AU)                 & \asemi        & D     \\
Incident Flux ($S_\oplus$)               & \sinc          & D
\enddata
\tablecomments{\\
A: Based on {\it Kepler} photometry.\\
%B: Based on the radial velocities.\\
B: Based on an analysis of the Palomar spectra.\\
C: Based on  stellar evolution tracks.\\
D: Based on Newton's version of Kepler's Third Law and total mass.
%F: Assumes Bond albedo = 0.1 and complete redistribution.\\
%G: 1 sigma upper limit
}
\end{deluxetable}

We determined \teff\ for \kepst\ following the method of \citet{Mann2013}, which we briefly summarize here. We compared our optical spectrum to a grid of PHOENIX BT-SETTL models\footnote{\href{https://phoenix.ens-lyon.fr/Grids/BT-Settl/CIFIST2011}{https://phoenix.ens-lyon.fr/Grids/BT-Settl/CIFIST2011}} \citep{Allard2011}. We masked out regions of the spectrum that are poorly reproduced by atmospheric models. The fit included six nuisance parameters to account for errors in wavelength and flux calibration and the offset between the blue and red arms of the spectrum. We derived an error on \teff\ based on the scatter in the model fits and a comparison between \teff\ values derived this way and those determined empirically from long-baseline optical interferometry \citep{Boyajian2012}. Our final \teff\ is 3240$\pm$61~K.

We combine our optical spectrum with the formula from \citet{Mann2013a} to determine the host star's [Fe/H]. \citet{Mann2013a} present empirical relations between the strength of atomic lines in visible and near infra-red M dwarf spectra and the metallicity of the host star, calibrated using a set of wide FGK+M dwarf binaries. Using the calibration for visible-wavelength lines we calculated [Fe/H] = -0.15$\pm$0.11. 

%\textbf{I.A. how was the stellar gravity calculated? Should we mention $L_* = 0.0063^{+0.0035}_{-0.0025}?$}

%{\bf S.H.: How does this calculation agree or relate to the estimates from the isochrone tracks?}

We calculated the stellar radius ($R_\star$) from our \teff\ and [Fe/H] using the relations from \citet{Mann2015b}, which are based on nearby single-stars with precise ($<5\%$) parallaxes. Our stellar radius errors account for both errors in \teff\ and [Fe/H], and the scatter in relations from \citet{Mann2015b}. We used these relations in conjunction with stellar evolution tracks to calculate a stellar radius of $R_\star=0.252\pm0.039 R_\sun$. We also used isochrone model fits to determine the stellar gravity \logg=$4.98\pm0.22$ as outlined in \citet{Teske2015} to arrive at a mass $M_\star=0.219\pm0.022 M_\sun$.

\section{Planet Propoerties}\label{sec:phot}

%\textbf{S.H.: Should we expect figure 3 to look like the bottom panel of figure 1? And if so, why do the data points in figure 1 go far below the transit curve when this does not occur for figure 3? Also, figure 3 does not look like it has enough data points.}

Photometry provided by the \ik project is based on apertures meant to maximize the S/N based on positions and apparent magnitudes from the KIC. However, it has previously been shown that for targets fainter than 15--16th mag in the Kepler bandpass that photometry based on a model of the point-spread function (PSF) is more precise \citep{Rappaport2014}. This form of photometry utilizes the Kepler pixel response function that the mission has archived at the Mikulski Archive for Space Telescopes (MAST) \citep{Bryson2010b}.

\begin{figure}
\begin{center}
\includegraphics[width=0.45\textwidth]{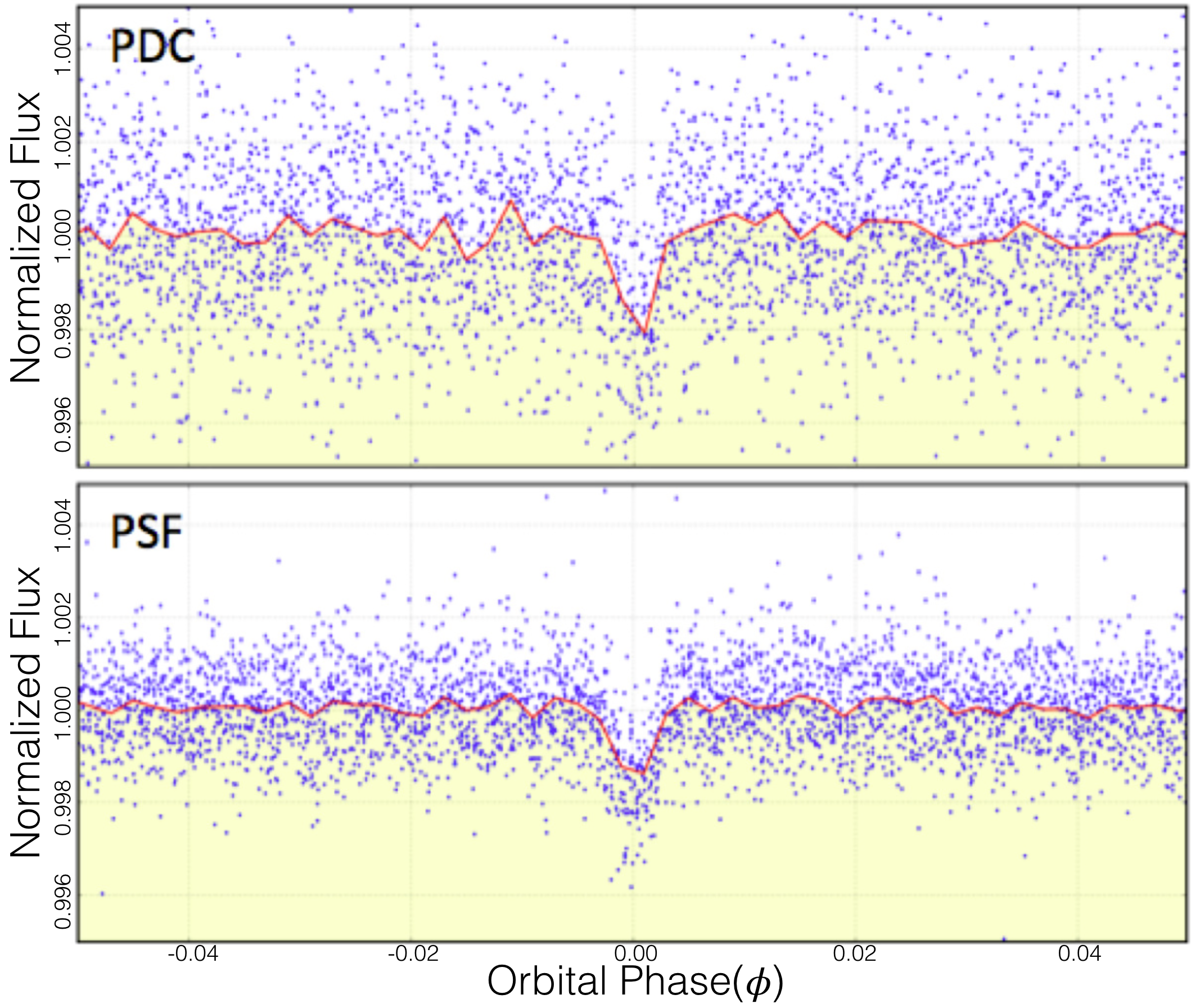}
\end{center}
\caption{The 10-quarter Kepler light curve collected from \kepst\ . The light curve has been extracted from calibrated pixels in 2 separate ways: all pixels within the photometric aperture defined by the Kepler pipeline are summed and then passed through the PDC algorithm (top) and a PSF model is fit to all pixels within the mask (bottom). Signals of astrophysical origin and systematics on timescales $> 1$ day have been removed and the light curves normalized. Both time series have been folded on a 8.68904 day orbital period with zero-phase corresponding to BJD 2,454,966.2406. Blue dots are individual observations and the red line is the same data median-averaged into 500 uniformly-sized phase bins.}
\label{fig:pdf}
\end{figure}

\begin{figure*}
\begin{center}
\includegraphics[width=0.95\textwidth]{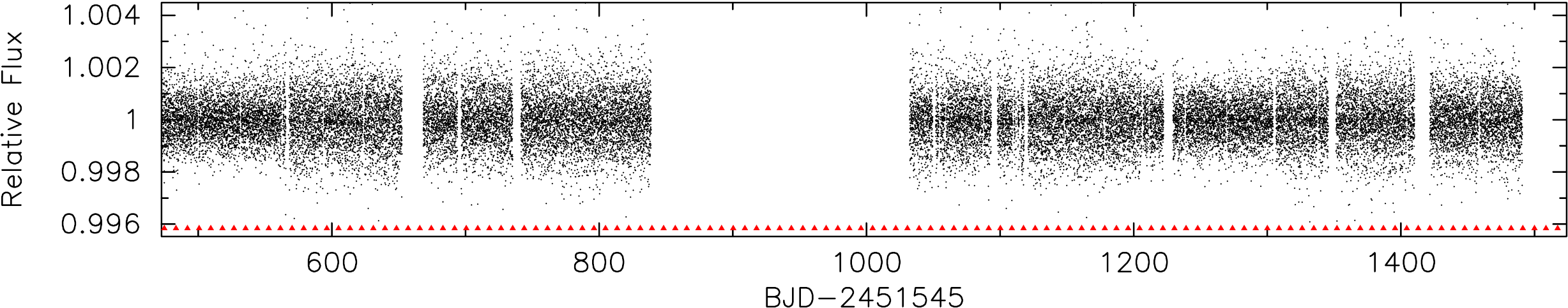}
\end{center}
\caption{The full Kepler time series data for \kepst\, created using PSF photometry. Transits occur every 8.7 days and are indicated by red triangles. These data cover Kepler observing Quarters 6 to 9 and 12 to 17. No data was collected during Quarters 10 and 11, the cause of the gap in the center.}
\label{fig:wholelc}
\end{figure*}

Utilizing the method described in \S3.1 of \citet{Rappaport2014}, we modeled \kepst\ and six stars near the target that had initial position and brightness values based on either the KIC or two different multicolor surveys of the Kepler field \citep{Everett2012,Greiss2012} using software provided by the Kepler Guest Observer Office \citep{Still2012}. The scatter integrated over the transit duration for \keppl\ of 1.04-hours, also known as the 1.04-hour Comined Differential Photometric Precision (CDPP) \citep{Christiansen2012} from the Kepler data, was 860 ppm whereas our PSF photometry had a scatter of 347 ppm, an increase in signal to noise of a factor of 2.5. This increase in photometric quality can be seen in Figure~\ref{fig:pdf}, where the upper panel is the Kepler pipeline derived photometry, summed and passed through a Presearch Data Conditioning (PDC) algorithm \citep{Smith2012,Stumpe2012,Stumpe2014}, and the lower panel was created using PSF photometry. Our PSF photometry for \kepst\ is utilized for the remainer of this paper for deriving planet properties of \keppl  (see Table \ref{tab:parameters}). The full time series of \kepst\ is shown in Figure~\ref{fig:wholelc}.

\begin{figure}
\begin{center}
\includegraphics[width=0.50\textwidth]{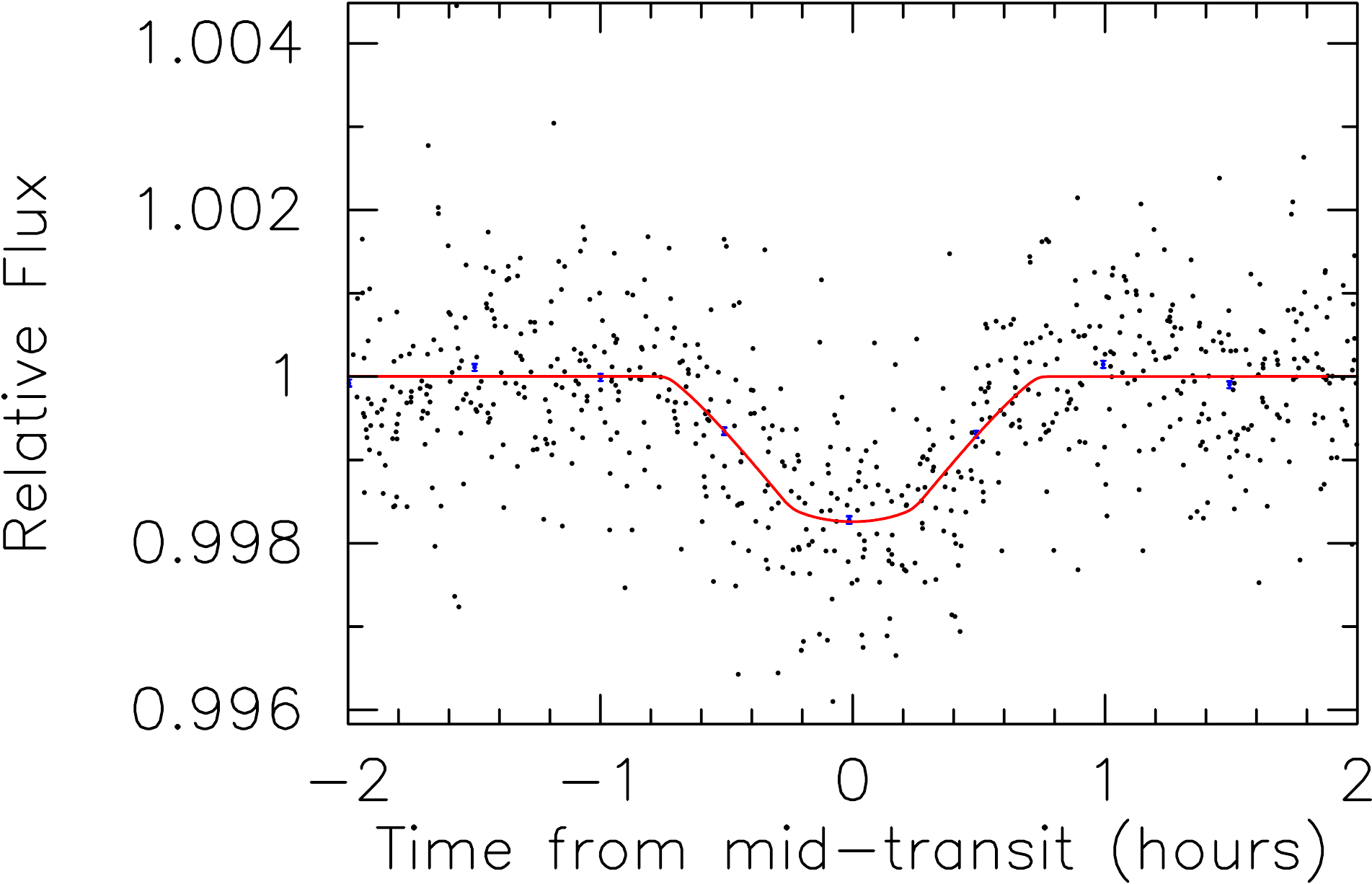}
\end{center}
\caption[Spectrum]{Phase-folded flux time series for \kepst. The flux time-series has been folded on the planet's orbital period. Each black point represents 1 observed datum, and the red curve shows a best fit transit curve.}
\label{fig:stransit}
\end{figure}

We fit a model to the observed transit of Kepler-XX to determine the properties of both the planet and host star.  We used the transit model of \citet{Rowe2014} which is described by a Keplerian orbit and transit based on the analytical description of \citet{Mandel2002} for quadratic limb-darkening.  The modelled parameters were the orbital period ($P$), time of first transit ($T_{0}$), ratio of the planet and star radius (\rprs), the impact parameter ($b$), and the mean stellar density (\rhostar). In addition to the model assumption that the mass of the planet is much less that the mass of the star, a circular orbit was adopted to perform our calculations.  The best fit parameters were found via least-squares analysis. The best fitting model is shown over-plotted on the phase-foleded transit data in Figure \ref{fig:stransit}.

To estimate the posterior distribution on each fitted model parameter, we used a Markov chain Monte Carlo (MCMC) approach similar to the procedure outlined in \citet{Ford2005}, but modified to better handle correlated variables as implemented in \citet{Rowe2014}.  The chain generation steps use a combination of Gibbs sampling or vectorized jump via random selection between the two methods. The latter method, vectorized jumps, uses a control set of model parameter sets and scale parameters as described in \citet{Gregory2011}. The adopted methods allow for efficient parameter space exploration even with highly correlated variables.  The generation of the chains was initially seeded with the best-fitting parameters found from the least-square fit.

We generated 10$^6$ Markov-chains, the first 20$\%$ of which were discarded as burn-in.  The remaining chains were combined into one continuous set and used to calculate the median, standard deviation and 1$\sigma$ bounds of the distribution centered on the median for each model parameter.  The transit and orbital parameters that were derived with the Markov chain include: orbital period ($P$), midtransit time ($E$), scaled semimajor axis (\adrs), scaled planet radius (\rprs), impact parameter ($b$) and orbital inclination ({{\it i}}). The orbit was assumed to be circular. We then used the Markov Chains to compute model-dependent measurements for the limb-darked transit depth at mid-transit, $\Delta F/F$=\tdep, and full transit duration, T$_{dur}$=\tdur.

\begin{figure}
\begin{center}
\includegraphics[width=0.45\textwidth]{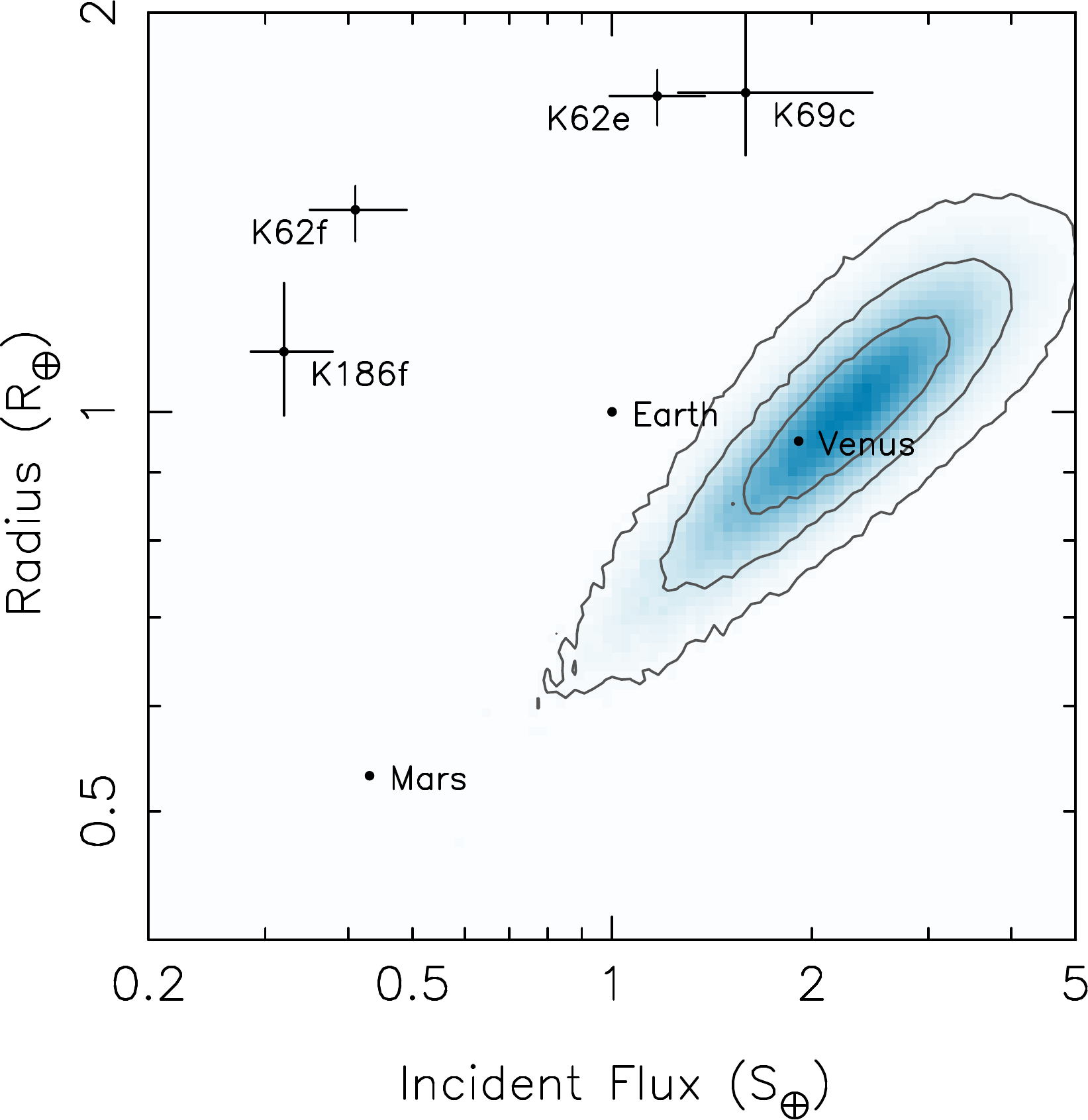}
\end{center}
\caption[Radius vs. Incident Flux]{The radius and incident flux for \keppl\ based on MCMC analysis.  The contours represent 1, 2, and 3$\sigma$ confidence levels.  Solar system objects Venus, Earth and Mars are also plots as well as a sample with 1$\sigma$ uncertainties of confirmed planets from \ikt: Kepler-186f \citep{Quintana2014}, Kepler-62e \citep{Borucki2013}, Kepler-62f \citep{Borucki2013}, Kepler-69c \citep{Barclay2013}.}
\label{fig:radsrad}
\end{figure}

%We next estimated the mass (\mstar = 0.24$\pm$0,04) and radius (\rstar = 0.247$\pm$0.032) of our cool M5V host star from stellar evolution tracks. 
We convolved our transit model parameters with the stellar parameters to compute the planetary radius, \rpl=\prad, orbital semimajor axis $a$=\asemi\ AU, and flux received by the planet relative to Earth $S$=\sinc. Our final results are presented in Table \ref{tab:parameters}. %Our values derived in \S\ref{sec:spec} agree with those derived from \ik photometry and stellar evolution tracks to within uncertainty limits and thus confirm the derived stellar parameters in Table 1.
Figure \ref{fig:radsrad} shows the range in incident flux and radius for \keppl\ from our MCMC analysis, showing that the planet has values consistent with Venus, even at the 1-$\sigma$ confidence level.

%The stellar model Markov-chain properties and generation is described in \S\ref{sec:spec}.

%Our adopted model parameters and 1$\sigma$ uncertainties are reported in Table 1.  

%To estimate the planetary radius, R$_{\rm p}$, and the incident flux received by the planet relative to the Earth, $S$, we convolved the transit model Markov-chain with the stellar model Markov-chain. These results are also shown in Table 1.  The stellar model Markov-chain properties and generation is described in \S\ref{sec:spec}.

% EQ: Note I moved up this section from end since it says virtually the same as above text
%\section{Planet Properties}\label{sec:planetprop}

\section{Validation}\label{sec:validation}
We performed a series of analyses and follow-up observations to eliminate the possibility of a false postive and validate \keppl\ as a planet. Although false detections due to noise are highly unlikely \citep{Jenkins2002}, it remains possible that our detected transit signal in the \kepst\ system is due to some other astrophysical source. Such false positive signals may be induced by background or foreground eclipsing binary systems, background or foreground transiting planet systems, or a planet transiting a bound companion to the target star. 

\begin{figure}
\begin{center}
\includegraphics[width=0.5\textwidth]{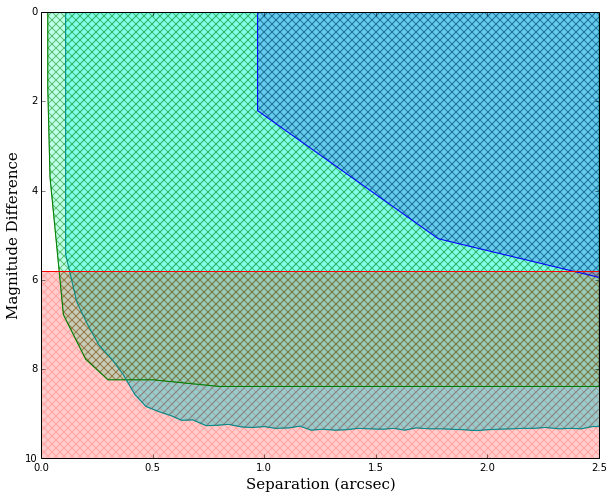}
\end{center}
\caption[exlusion zones]{Exclusion zones for \kepst\ in which a false positive source cannot reside. All curves are within $3-\sigma$ certainty. Regions eliminated from \ik transit data are shown in red, from UKIRT imaging in blue, from speckle data in green, and from AO data in cyan. We cannot rule out the possibility of a false positive source residing in the white regions of this figure and thus account for it in our false positive analysis.}
\label{fig:exclusion}
\end{figure}

Figure~\ref{fig:exclusion} shows the $3$-$\sigma$ regions of stellar magnitude and separation parameter space that we eliminate as potential locations of a false positive source. The procedures outlined in this section were conducted to assure that no such sources were detected in these regions. We conclude from our observations that the existence of a false positive source is highly unlikely, thus allowing our signal to be interpreted as a planet transit associated with the \kepst\ system.

%\subsection{Constraining Potential False Positive-Hosting Parameter Space}

\begin{figure}
\begin{center}
\includegraphics[width=0.45\textwidth]{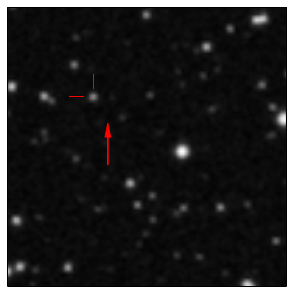}
\caption{Palomar Observatory Sky Survey (POSSII) image of the region near \kepst\ illustrating the lack of background sources. The image is 2.25 $\times$ 2.25 arcmin on a side with North up and east to the left. The red arrow indicates the location in which \kepst\ currently resides, and the two lines indicate the location of \kepst\ when the image was taken. No background objects are detected to a depth of $\sim$22.5 magnitudes on or near the star's current location.}
\end{center}
\label{fig:palomar}
\end{figure}

To assess the probability of a false positive, we systematically eliminated various regions of parameter space in which a false positive-inducing source can exist. We began by using the proper motion of the target star to eliminate the chances of a confounding background source. \kepst\ is a high proper motion star (0.157$\arcsec$ yr$^{-1}$) and thus resides in a slightly different location in the sky today than it did just decades ago \citep{Lepine2005}. We can examine its current location in older images for any possible background stars that may be the source of a false positive signal. Examination of images from the STScI Digitized Sky Survey taken on 06 Sept 1991 as part of the POSSII-F Sky Survey \footnote{\href{http://stdatu.stsci.edu/cgi-bin/dss_form}{http://stdatu.stsci.edu/cgi-bin/dss$\_$form}} are shown in Figure 7. No background objects are detected at the current location of \kepst\ in this image. The POSSII-F image has a plate limit of R$\sim$22.5$\pm$0.4 \citep{Reid1991}, meaning that a confounding background source would have to be more than $\sim$5 magnitudes fainter than \kepst. Such a source is likely not bright enough to produce the observed transit, as we show in the next paragraph.

We analyzed the \ik transit data to place constraints on the magnitude of the transiting object. According to the MCMC transit analysis outlined in \S\ref{sec:phot}, the detected transit in question has a measured depth of \tdep . In the case that the transit was induced by an eclipsing binary system, we can calculate the maximum possible magnitude of the transiting object by assuming the system undergoes total eclipes \citep{Chaplin2013}. Under this inference such a depth requires that the source be at most 5.8 magnitudes fainter than the target star in order to fit our transit model. We can thus rule out all nearby stars that are more than 5.8 magnitudes fainter than \kepst\ as possible transit sources. The red exclusion zone in Figure~\ref{fig:exclusion} indicates the region of parameter space in which these stars would reside. Any star not seen with POSSII-F would exist in this region, thus allowing us to completely eliminate the possibility of a background transit source.

%Additionally, any false positive source physically bound to KOI-3138 must be a brown dwarf in order to satisfy this magnitude constraint, as discussed further in \S\ref{sec:binary}.
%After eliminating the chances of a background false positive source,
We were then left with ruling out the possbility that \keppl\ orbits a bound companion to \kepst\ that cannot be detected in a single spectrum \citep{Teske2015} or resolved by \ik or POSSII images, in which case our transiting planet would be larger than the size derived in \S\ref{sec:phot}. Our next step thus involved inspection of seeing-limited follow-up images of \kepst\ to reveal any unresolved companions that could host a transiting planet. The first of these images were taken in the J-band by the UK Infrared Telescope (UKIRT). The UKIRT images reveal several stars that fall within a few arcseconds of our target star. All of the resolved stars, given their distance to \kepst, were ruled out as possible transit sources because they did not induce a correlated shift in the photo-center of \kepst\ over time, a common characteristic of eclipsing binaries \citep{Bryson2013}. Any false positive sources further from our target star would have been seen in our UKIRT images and can thus be ruled out, as indicated by the blue region in Figure~\ref{fig:exclusion}. There is, however, a seeing limit to images taken with UKIRT of about 0.9 arcseconds. We cannot eliminate areas within this seeing limit and therefore rely on alternative methods to explore regions unaccounted for in UKIRT.

High resolution speckle images of \kepst\ were obtained on 15 July 2015 using the DSSI imaging camera mounted on the 8-m Gemini-N telescope. Observations with DSSI are taken simultaneously in two filters. This observation used a 692 nm center-wavelength filter with a 40 nm width and an 880 nm center-wavelength filter of width 50 nm. The seeing was superb, near 0.4-0.5 arcsec throughout with a total of 20 minutes spent collecting 60 msec frames on \kepst. Details of the observational procedure and data reduction techniques are given in \citet{Horch2012}.

\begin{figure}
\begin{center}
\includegraphics[width=0.45\textwidth]{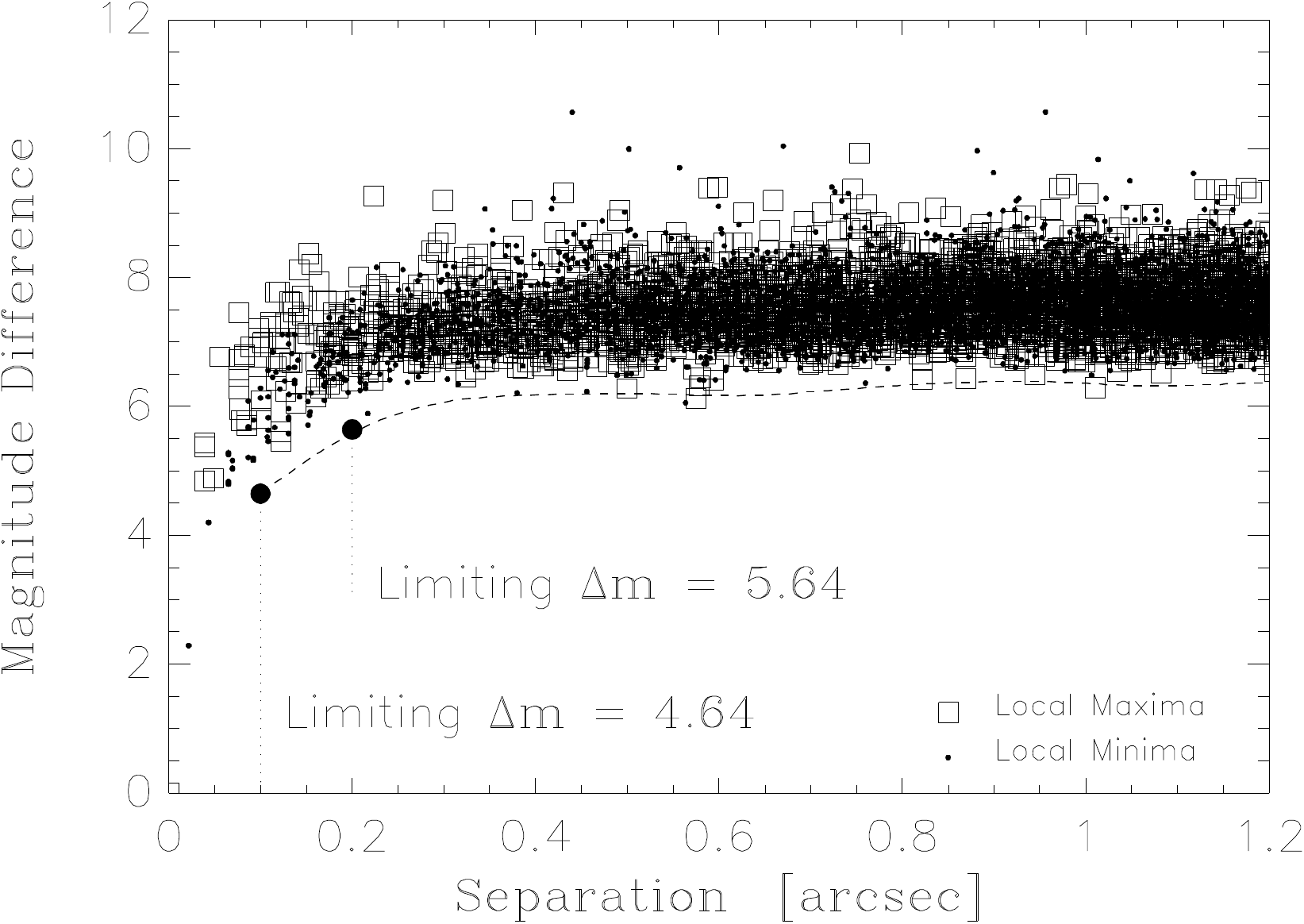}\\
\includegraphics[width=0.45\textwidth]{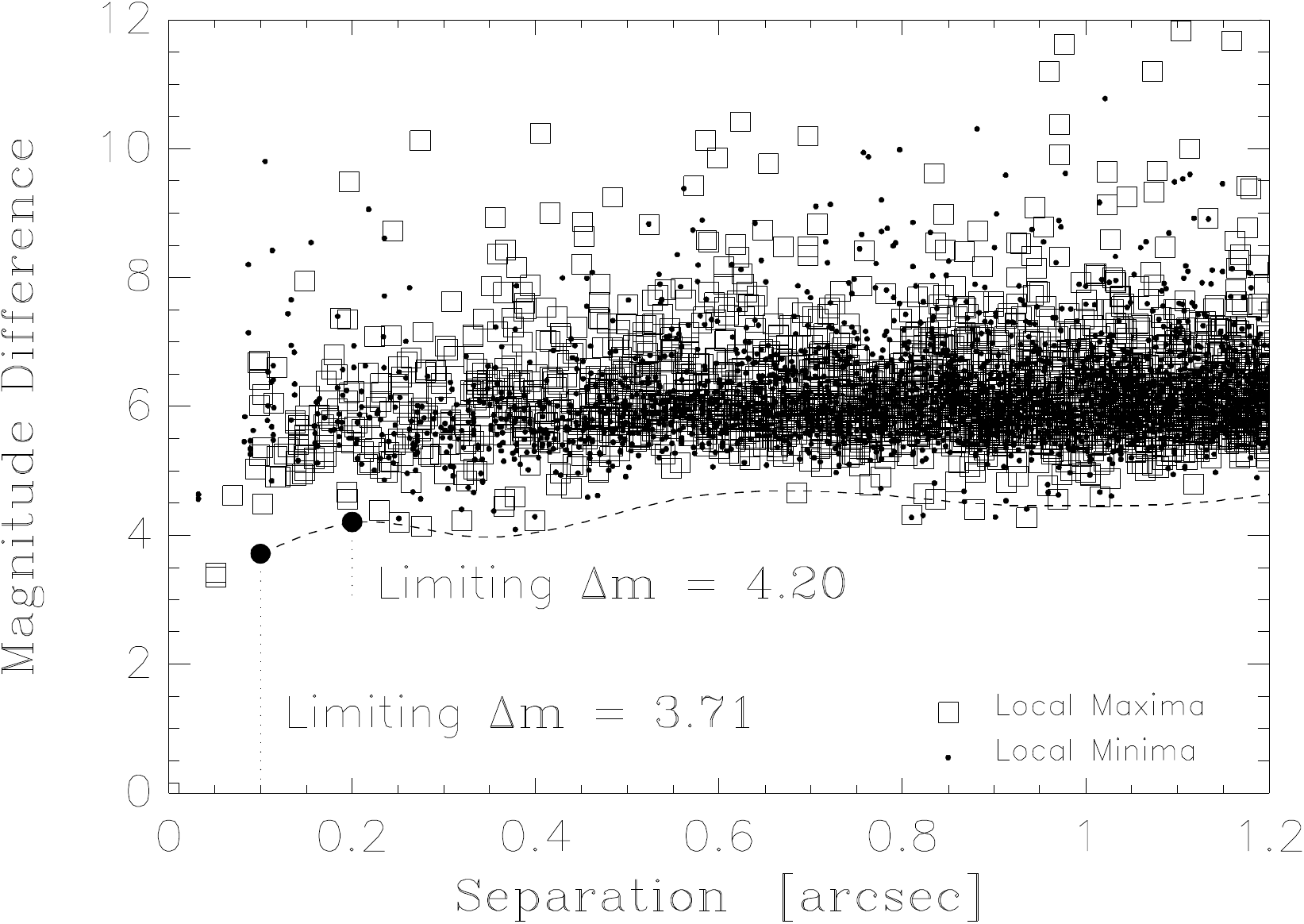}
\end{center}
\caption{Detection limit analysis for the observation of \kepst. In both plots, the dashed line represents the formal 5$\sigma$ limiting magnitude as a function of separation, as described in the text. The result in the 692 nm filter 880 nm filter are shown in the top and bottom panels respectively.}
\label{fig:detlimits}
\end{figure}

Figure~\ref{fig:detlimits} shows the results from our two high-resolution speckle images, where the upper limit in magnitude as a function of separation is given by the dashed line. \kepst\ is a faint star for speckle observations, but both images nonetheless find at high significance that no companion star exists. The image at 692 nm reveals that at $5$-$\sigma$ no companion within a magnitude difference $\Delta Kp\sim5\:{\rm  mag}$ exists down to a spatial radius of 0.022 arc sec. The 880 nm image is somewhat poorer, finding that no companion exists at $5$-$\sigma$ to a $\Delta Kp$ of near 4 magnitudes into 0.027 arc sec, as indicated by the green area of Figure~\ref{fig:exclusion}. 

%\textbf{EQ: Are we supposed to be showing speckle images too, or is this describing the area under the curves in Fig 7?}

\begin{figure*}
\begin{center}
\includegraphics[width=0.9\textwidth]{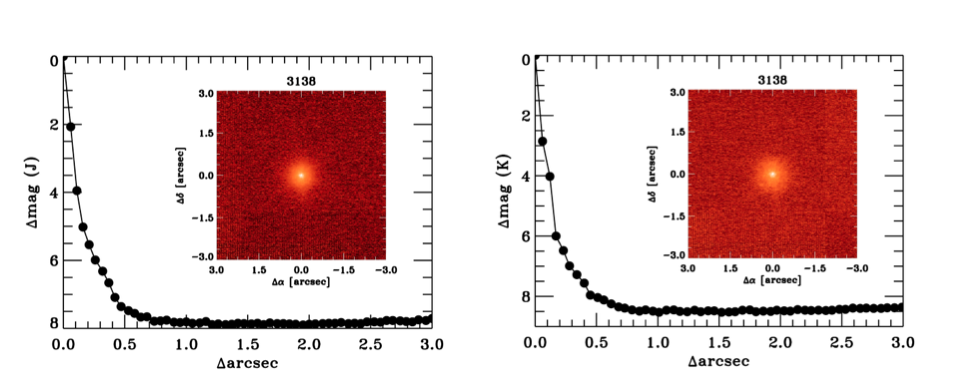}
\caption{3$\sigma$ sensitivity limits on adaptive optics images of \kepst\ from the Keck II Telescope obtained on July 21, 2015.The left panel shows the J-band image and sensitivity curve and the right panel shows the same for the K-band filter. Neither image reveals a possible bound companion. }
\end{center}
\label{fig:ao}
\end{figure*}

%Results from these observations, when converted from $5$-$\sigma$ to $3$-$\sigma$, allowed us to eliminate all plausible false positive inducing stars in the green area of Figure 6.

Near-infrared adaptive optics (AO) images of \kepst\ were taken in the J (1.248 $\mu$m) and K (2.15 band $\mu$m) band filters. The images were taken using the NIRC2 imager on the Keck II Telescope on UT 2015, June 21 and are shown in Figure 9. Also shown are the sensitivity limits of the NIRC2 imager as a function of radial distance from the host star. As illustrated in the figures, no companions can be seen with adaptive optics imaging in any region within $\tilde{0.1}-3''$ in the J and K-band filters to within a magnitude difference of 7.48 at 0.5 arcsec in the J-band. This data is also converted to Kepler magnitudes according to methods outlined in \citep{Howell2012}. The cyan region of Figure~\ref{fig:exclusion} represents areas of parameter space that can be eliminated as companion-hosting, according to our AO observations.
% and overlaid with the others in Figure 6 

Given the data presented in Figure~\ref{fig:exclusion}, a false positive-inducing bound companion to \kepst\ can only exist in the parameter space indicated by the small white region, $\sim$5 mag fainter than the target. Additionally, in order to remain consistent with our \ik transit, a companion transit source must be a maximum of 5.8 magnitudes fainter than our target star. From this information we assess the probability that the signal was induced by a planet transit around a star physically associated with \kepst. We first determined the mass and radius of the faintest possible binary companion to \kepst\ using COND03 isochrone models \citep{Baraffe2003}. We computed the mass for a star 5.8 magnitudes fainter than \kepst\ and found a lower limit of 0.05 \msun\ and a corresponding radius of 0.093 \rsun.

According to the isochrone models, a companion with a mass of 0.05--0.07 \msun\ would be between 4.028 and 5.390 magnitudes fainter than our host star, placing it near the M star-brown dwarf boundary. Fitting these star parameters to transit curves from \ik data reveal that a planet transiting around a companion would have a radius of $\approx$2.8--4.7 \rearth. Such a scenario would be consistent with our observations, however, the prospect of the system existing in the first place is highly unlikely given the low occurence rates ($\leq0.15$ star$^{-1}$) of large planets around cool stars \citep{Berta2013}.

Given the low likelihood of a nearby, bound false positive source, in conjunction with our non-detection of a confounding background binary system, we verify that the transit signal around \kepst\ is due to a planet, \keppl\, orbiting the system.

\section{Discussion}\label{sec:discussion}
We present the discovery and planetary confirmation of \keppl, an approximately Earth-sized planet in a 9-day orbit around a nearby M5V star receiving an incident flux of \sinc\ relative to Earth. We cannot derive a mass estimate for the planet from photometric data alone and currently don't have constraints from transit timing analysis or radial velocity measurement. Additionally \keppl\ is too small to induce a detectable ``wobble" in its host star which could provide future constraints on its mass. We therefore make no conclusions about mass or composition in this paper. Planets with sizes comparable to Earth, however, have a high likelihood of being rocky \citep{Rogers2015}.

\keppl\ is comparable in size and host star to Kepler-186f, an Earth-sized exoplanet discovered to orbit in the habitable zone of an M dwarf \citep{Quintana2014}. Both planets orbit cool stars and thus exist in systems that are significantly different to that of the terrestrial planets in our Solar System. Kepler-186f orbits a star that is about half a solar radius with a 130 day period, while \keppl\ orbits with an 8.7 day period around a star that is about a quarter the size of our Sun. Because of this, the two planets may be more prone to effects of host star variability such as flares and coronal mass ejections than Earth and Venus. They also receive comparatively low-energy radiation due a shift in the spectral energy distribution for M Dwarfs relative to the Sun. Furthermore, because \keppl\ and Kepler-186f orbit much closer in than those of Venus and Earth, they may be subject to larger tidal effects from their host star. These effects may include tidal heating, synchronous rotation, and tidal locking, which can produce a significant effects on the planets' seasons and geologic activity. 

Regardless, due to their size and incident flux, planets like \keppl\ and Kepler-186f are good candidates for Earth- and Venus-analog studies. In terms of insolation, \keppl\ is too hot to reside within its star's habitable zone and instead is in the so-called ``Venus Zone", a Venusian analog to the habitable zone as described in \citet{Kane2014}. The discovery of \keppl\ thus highlights the relatively high abundance of terrestrial planets that may have runaway greenhouse surface environments, lending itself to future studies surrounding exoplanet atmospheres and habitability.

%However, the size of the host stars, and the proximity of the planets to their star, differ. Kepler-186 is about half a solar radius and Kepler-186f orbits with a 130 day period. /\kepst\/ is about a quarter the size of our Sun, so given the lower luminosity its habitable zone will be much closer in compared to the Kepler-186 case. While \keppl\ orbits with an 8.7 day period, it receives much more light from its host star than Kepler-186f 

Distinguishing between Earth and Venus analogs is becoming especially important as the ongiong K2 mission, like \ik, discovers and studies more and more Earth-sized, near-habitable zone planets \citep{Demory2016}. It will also remain important for this same reason as the upcoming Transiting Exoplanet Survey Satellite (TESS) mission preferentially detects and observes planets that are close to their host star \citep{Ricker2014}. These discoveries will lend themselves well to observations with the James Webb Space Telescope (JWST), which will have the potential to probe the atmospheres of planets such as \keppl\ and ultimately constrain their habitability \citep{Greene2016}. Such observations can help us understand the correlation between insolation flux (\sinc S$_\oplus$ for \keppl, easily observable for missions like \ik) and habitability. We can also learn about other factors, like additional undetected planets in the system or tidal effects due to proximity to the host star, that may further contribute to a Venus-like climate \citep{Barnes2013}.

%the proximity of the planet to the host star that may induce tidal effects that would further contribute to a Venus-like climate \citep{Barnes2013}. 

Similar to planets that lie within a star's habitable zone, confirmation of surface conditions of \keppl\ would require a detailed spectroscopic analysis of the atmosphere. Facilities capable of extracting such measurements for this planet are unlikely to be available in the near-term. The detection of Venus analog atmospheres via methods listed above presents a significant challange due to the opacity of the Venusian atmosphere, though there are distinguishing features at high altitudes including carbon dioxide absorption combined with an upper haze layer with sulphuric acid \citep{Barstow2016,Ehrenreich2012}. A cloud-dominated atmosphere also produces large scattering and reflection effects that translate into a relatively high geometric albedo, producing another source of evidence linking the atmosphere to a runaway greenhouse \citep{Kane2013}. Adopted a Venusian geometric albedo of 0.65, we calculate a flux ratio amplitude between the planet and the host star of $5.2 \times 10^{-7}$. For comparison, Venus in our Solar System produces a phase amplitude of $2.0 \times 10^{-9}$ \citep{Kane&Gelino2013}. The predicted phase variation amplitude of the planet is beneath the noise threshold of the Kepler photometry, but could be examined as a diagnostic from follow-up observations of similar Venus analog candidates.

Further constraints on the stellar parameters of \kepst\ are needed to increase the accuracy of our predicted planet properties. Fortunately, the GAIA spacecraft is a space-based telescope capable of measuring distances to nearby systems like \kepst\ by taking precise parallax measurements \citep{Stassun2016}. Such distance measurements will help place constraints on the luminosity of \kepst\ and thereby further increase the accuracy of the star and planet parameter calculations summarized in Table 1. 

The discovery of \keppl\ is part of a larger movement towards confirmation and characterization of a variety of Earth-sized exoplanets with the ultimate goal of understanding what factors place constraints on habitabiilty. Most of these planets have orbital periods measured to high precision, allowing us to calculate the flux received by the planet from its host star. As a result, determining the correlation between incident flux and atmospheric compositions would be highly useful in assessing the habitability of known exoplanets. More specifically, determining the compositions and atmospheres of planets like \keppl\ and Kepler-186f, two planets that together span a wide range of distances within the habitable zones of M Dwarfs, will be useful in understanding the nature of habitable zone boundaries for such star types. Future missions like K2, TESS, and JWST as described above, will make these studies possible and therefore lend themselves to a better understanding of conditions required for exoplanet habitability. 

\acknowledgments

J.F.R. acknowledges NASA grants NNX12AD21G and NNX14AB82G issued through the Kepler Participating Scientist Program. The research presented in this paper includes data collected by the Kepler Mission. This research has made use of the NASA Exoplanet Archive, which is operated by the California Institute of Technology, under contract with the National Aeronautics and Space Administration under the Exoplanet Exploration Program. Data from the Mikulski Archive for Space Telescopes (MAST) was used for this research as well. Data reduction was done using Image Reduction and Analysis Facility (IRAF) routines. E.V.Q was supported by an appointment to a NASA Postdoctoral Program Senior Fellowship at NASA Ames Research Center, administered by Universities Space Research Association under contract with NASA. BB acknowledges financial support from the European Commission in the form of a Marie Curie International Outgoing Fellowship (PIOF-GA-2013- 629435). Some of the data presented herein were obtained at the W.M. Keck Observatory, which is operated as a scientific partnership among the California Institute of Technology, the University of California and the National Aeronautics and Space Administration. The Observatory was made possible by the generous financial support of the W.M. Keck Foundation. The authors wish to recognize and acknowledge the very significant cultural role and reverence that the summit of Mauna Kea has always had within the indigenous Hawaiian community.  We are most fortunate to have the opportunity to conduct observations from this mountain.

\bibliography{AstroRefs.bib}

\end{document}